\documentclass{article}

\usepackage{moresize}
\usepackage{arxiv}

\usepackage[utf8]{inputenc} 
\usepackage[T1]{fontenc}    
\usepackage{hyperref}       
\usepackage{url}            
\usepackage{booktabs}       
\usepackage{amsfonts}       
\usepackage{nicefrac}       
\usepackage{microtype}      
\usepackage{subfigure}
\usepackage{graphicx}
\usepackage{doi}

\title{Exploring the Role of Randomization on Belief
Rigidity in Online Social Networks}

\author{ Adiba Mahbub Proma\\
	Department of Computer Science\\
	University of Rochester\\
        Rochester, NY \\
	\texttt{aproma@cs.rochester.edu} \\
	\And
	{Neeley Pate} \\
	Department of Computer Science\\
	University of Rochester\\
        Rochester, NY \\
	\texttt{npate@ur.rochester.edu} \\
	\And
	Raiyan Abdul Baten \\
	Department of Computer Science and Engineering\\
	University of South Florida\\
        Tampa, FL \\
	\texttt{rbaten@usf.edu} \\
	\And
	{Sifeng Chen} \\
	Department of Computer Science\\
	University of Rochester\\
        Rochester, NY \\
	\texttt{schen133@u.rochester.edu}\\
        \And 
        James Druckman\\
	Department of Political Science\\
	University of Rochester\\
        Rochester, NY \\
	\texttt{jdruckma@ur.rochester.edu} \\
	\And
	{Gourab Ghoshal} \\
	Department of Physics and Astronomy\\
	University of Rochester\\
        Rochester, NY \\
	\texttt{gghoshal@pas.rochester.edu} \\
        \And 
        {Ehsan Hoque} \\
	Department of Computer Science\\
	University of Rochester\\
        Rochester, NY \\
	\texttt{mehoque@cs.rochester.edu}
}

\renewcommand{\shorttitle}{Exploring the Role of Randomization on Belief Rigidity in Online Social Networks}

\hypersetup{
pdftitle={Exploring the Role of Randomization on Belief
Rigidity in Online Social Networks},
pdfauthor={Adiba Mahbub Proma, Neeley Pate, Raiyan Abdul Baten, Sifeng Chen, James Druckman, Gourab Ghoshal, Ehsan Hoque},
pdfkeywords={Belief Rigidity, Online Social Networks, Recommendation, Experimental Framework},
}

\begin{document}
\maketitle

\begin{abstract}
People often stick to their existing beliefs, ignoring contradicting evidence or only interacting with those who reinforce their views. Social media platforms often facilitate such tendencies of homophily and echo-chambers as they promote highly personalized content to maximize user engagement. However, increased belief rigidity can negatively affect real-world policy decisions such as leading to climate change inaction and increased vaccine hesitancy. To understand and effectively tackle belief rigidity on online social networks, designing and evaluating various intervention strategies is crucial, and increasing randomization in the network can be considered one such intervention. In this paper, we empirically quantify the effects of a randomized social network structure on belief rigidity, specifically examining the potential benefits of introducing randomness into the network. We show that individuals' beliefs are positively influenced by peer opinions, regardless of whether those opinions are similar to or differ from their own by passively sensing belief rigidity through our experimental framework. Moreover, people incorporate a slightly higher variety of different peers (based on their opinions) into their networks when the recommendation algorithm provides them with diverse content, compared to when it provides them with similar content. Our results indicate that in some cases, there might be benefits to randomization, providing empirical evidence that a more randomized network could be a feasible way of helping people get out of their echo-chambers. Our findings have broader implications in computing and platform design of social media, and can help combat overly rigid beliefs in online social networks. 
\end{abstract}


\section{Introduction}
Humans tend to stubbornly stick to their beliefs and pre-existing opinions, often dismissing information that contradicts their beliefs to prevent cognitive dissonance \cite{scheffer2022belief, cotton2013cognitive}. Belief rigidity can be considered as how likely an individual is to change their stance in light of new information/opinions \cite{sharot2011unrealistic}. While rigid beliefs can ensure consistency in behavior \cite{scheffer2022belief}, harmful rigid beliefs can often lead to societal problems. For example, strong beliefs in vaccine conspiracy theories have been shown to increase vaccine hesitancy \cite{djordjevic2021links}, making it harder to achieve herd immunity during the COVID-19 pandemic \cite{cascini2021attitudes}. Moreover, people tend to seek others with shared beliefs, confirming their own biases through homophilic tendencies, and forming echo-chambers \cite{mcpherson2001birds, cinelli2021echo}. This is especially prevalent in social media platforms, which are generally designed to optimize for user engagement \cite{haroon2023auditing}. Social media platforms suggest similar content \cite{bakshy2015exposure, schmidt2017anatomy, cinelli2021echo}, thus limiting exposure to diverse perspectives \cite{cinelli2021echo, spohr2017fake}. Researchers found that the median Facebook user received most of their content from like-minded sources— 50.4\% compared to 14.7\% from cross-cutting sources \cite{nyhan2023like}. Mutual reinforcement among like-minded peers can result in increasingly stubborn beliefs over time, leading to polarization and increased susceptibility to accept and spread misinformation \cite{spohr2017fake, treen2020online}. This can have real-life implications on what policies people support, who they vote for, and how they perceive others in society. For instance, political inaction for climate change has often been attributed to partisan echo-chambers formed to challenge the consequence of climate change \cite{jasny2015empirical}. It is, therefore, essential to study belief rigidity within the context of online social networks and design interventions to reduce belief rigidity. 

In this paper, we aim to quantify the effects of a randomized social network structure on belief rigidity, specifically examining the potential benefits of introducing randomness into network connections. We consider increasing randomization as an intervention aimed to increase exposure to diverse perspectives. Our experimental framework enables passive sensing of private belief rigidity, thus allowing us to better understand individual belief rigidity. By introducing more randomization into the network, our study explores whether deviating from algorithmically curated personalized content can encourage individuals to consider a broader range of perspectives. Theoretical studies on opinion dynamics in social networks suggest that it is essential to consider the combined effect of human social influence and algorithmic decisions to understand how opinions are affected in online social networks \cite{santos2021link}. While our study focuses on belief rigidity, we can still draw from this literature on opinion dynamics. Extending upon the idea that both social influence and algorithmic choices have an impact, our randomization condition is designed to increase randomization in both human social influence aspect and in the recommendation algorithms. Our research questions are:

\begin{itemize}
    \item \textbf{RQ1:} How does exposure to a more diverse spectrum of opinions influence belief rigidity?  
    \item \textbf{RQ2:} Can a more randomized recommendation convince people to be more open to incorporating diverse views in their networks? 
\end{itemize}

Typically, social media analysis has been used to understand social signals on social platforms, including impact of similar content \cite{nyhan2023like}, impact of recommendation algorithms \cite{guess2023social}, and interaction with political content \cite{allcott2020welfare, bail2018exposure}. Different experiments have also been conducted on social media platforms to understand polarization \cite{nyhan2023like, guess2023social, asimovic2021testing}. However, studying change in beliefs through social media analysis and experiments is difficult because beliefs are private. Data from social media can only make an estimate about change in belief since online behavior may differ compared to people's private beliefs \cite{suler2004online, lieberman2020two}. For example, people may appear more stubborn in their stance compared to in real life to prove a point to others \cite{suler2004online}. Social media analysis on Reddit's subreddit r\textbackslash ChangeMyView sheds some light on how individuals change opinions \cite{tan2016winning}. However, the individuals engaged in this subreddit are \textit{open} to changing their stance, which is usually not the case. Therefore, we take an experimental approach to answer our research questions. Our experimental framework can help measure \textit{private belief} since participants have no other motive to change their beliefs unless they really want to.

\begin{figure*}
    \centering
    \includegraphics[width=.75\linewidth]{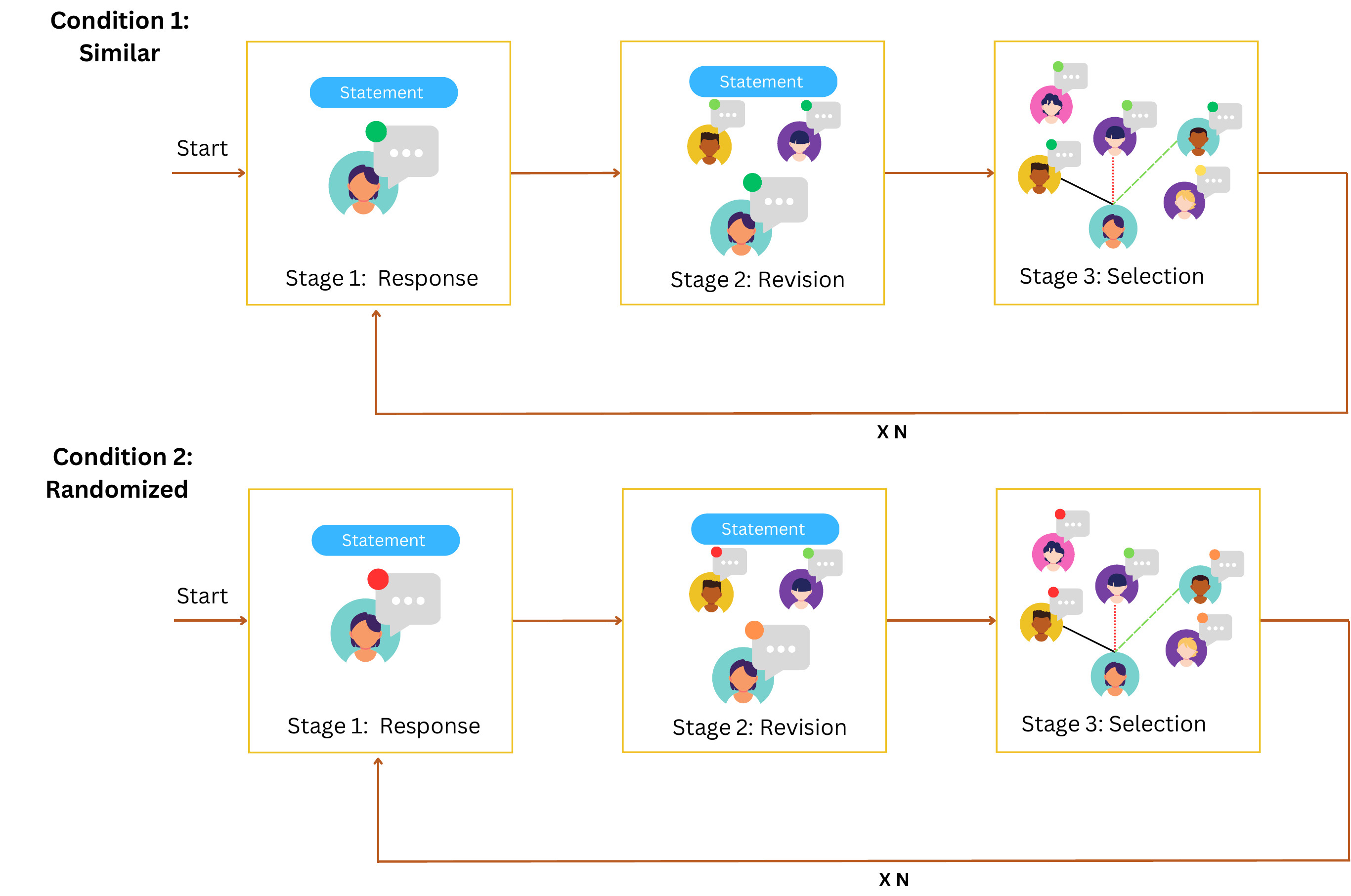}
    \caption{\small{In this figure, we describe the experimental framework, along with examples to understand the flow of the experiment. Condition 1 is designed to show similar users, while condition 2 is designed to show random users. Participants start by agreeing or disagreeing to a statement on a 7 point Likert Scale in stage 1. We denote a participant rating agreeably to the statement as green and rating disagreeably as red, while orange is neutral. In stage 2, for condition 1, when a participant rates the statement as green, they mostly see others who have rated it green or orange. A similar scenario would occur if they participant rated the statement red, as they would mostly see other reds or oranges. In condition 2, however, participants see a wide array of ratings. An example is provided in this figure, where even though the participant disagreed (denoted by red), they see users with red, orange and green ratings. Stage 3 is designed for network restructuring as participants can follow (denoted by green lines) and unfollow (denoted by red lines) people or stick to their previous choices (denoted by black lines).}}
    \label{fig:experimentalCondition}
\end{figure*}

Participants are given prompts which they rate on a 7-point Likert scale, and explain the reason for their rating. Next, they are shown the answers of other participants and asked if they would like to change their rating. This change in rating represents change in belief, and can be used to understand social influence. Participants are then asked to select three participants among those recommended whose responses they would like to see in the next round. Analyzing participant selections help in understanding the role of recommendation algorithms. Our experimental design consists of two conditions. For the first condition (Condition 1, C1: similar condition), we emulate traditional social media networks by recommending similar users and showing users that participants select themselves. In the second condition (Condition 2, C2: randomized condition), we increase randomization by recommending randomly, and only showing 1/3 of who participants selected (the rest is randomized) (Figure~\ref{fig:experimentalCondition}). 


Our analysis shows that irrespective of whether peer opinions are similar or diverse compared to one's own, they influence individuals' beliefs. Moreover, given a choice and access to a wide variety of viewpoints by the recommendation algorithm, individuals tend to include \textit{some} of those diverse viewpoints into their networks, but generally tend to prefer similar views  as themselves. In both conditions, participants followed people with similar views than not, but the effect was stronger in the similar condition group. Therefore, our results suggest that increased randomization can help individuals be slightly more open to differing views, while also highlighting people’s homophilic tendency. Our results have broader implications in computing and platform design, since most social media platforms tend to offer highly personalized content, thus facilitating people’s homophilic tendencies. While personalization has its merits, it can have unintended negative consequences on belief rigidity and polarization if not implemented carefully. Our findings provide empirical evidence that a more randomized network might be better suited to positively impact belief rigidity, and could help reduce echo chambers and polarization.

\section{Results}
To answer our research questions, we compare two conditions - Condition 1 (C1: similar condition) is designed to prioritize similar users in the network, while Condition 2 (C2: randomized condition) is designed to increase randomization in the network. This means that an individual in C1 is more likely to encounter perspectives closely aligned with their own, compared to a participant in C2, who would be presented with a wider array of differing opinions/views. We conducted five batches of each of the conditions, recruiting a total of 163 participants using Prolific Academia (C1: 82, C2: 81). Participants must all join at the same time online and the experiment approximately takes 60 minutes. Note that all correlations calculated are Pearson's correlation, $\rho$.

\subsection*{Study design}
Our study consists of five rounds; in each round, participants are given one prompt regarding climate change actions. Each round has three stages - response, revision, and selection (Figure~\ref{fig:experimentalCondition}). In the response stage (stage 1), participants see a prompt regarding spending money to mitigate climate change, and they are asked to rate the statement on a 7-point Likert scale, and provide reasoning for their rating. The 7-point Likert scale ranges from ``strongly disagree" to ``strongly agree". For the rest of the paper, we refer to each rating and its accompanying explanation as a ``response entry". 

In the revision stage (stage 2), participants are shown the response entries of selected participants (referred to as ``peers"), and are given the option to change their rating - we will refer to this as the ``updated rating". Note that participants are provided with their prior rating from stage 1 (response stage), thus ensuring that their change in rating is indeed intentional. For the first round of the similar condition (C1), we assign participants similar peers based on their ratings, and for the following rounds, participants see the response entries of peers they selected in stage 3 (selection stage) in the immediate previous round. For the first round of the randomized condition (C2), participants are randomly assigned three peers, and in the following rounds, they only see one of the peers they selected in stage 3 (selection stage). The other two peers are randomly assigned from those they did not select. 

In the selection stage (stage 3), participants are recommended response entries of other participants, and they also see the response entries of participants they saw in stage 2. For C1, participants are recommended similar response entries, while for C2, they are recommended randomly. Participants must select (i.e., ``follow") three peers whose answers they would like to see in stage 2 of the next round. They can stick to previous connections or pick new ones (or do a mix of both). However, as mentioned previously, only C1 participants can see the response entries of those they followed in stage 2 of the next round. C2 participants only get one of the peers they followed, and the other two are randomly selected.


\subsection*{Peer opinions influence belief, even when they are different from one's own}

\begin{figure*}
\centering
\begin{subfigure}
  \centering
  \includegraphics[width=.35\linewidth]{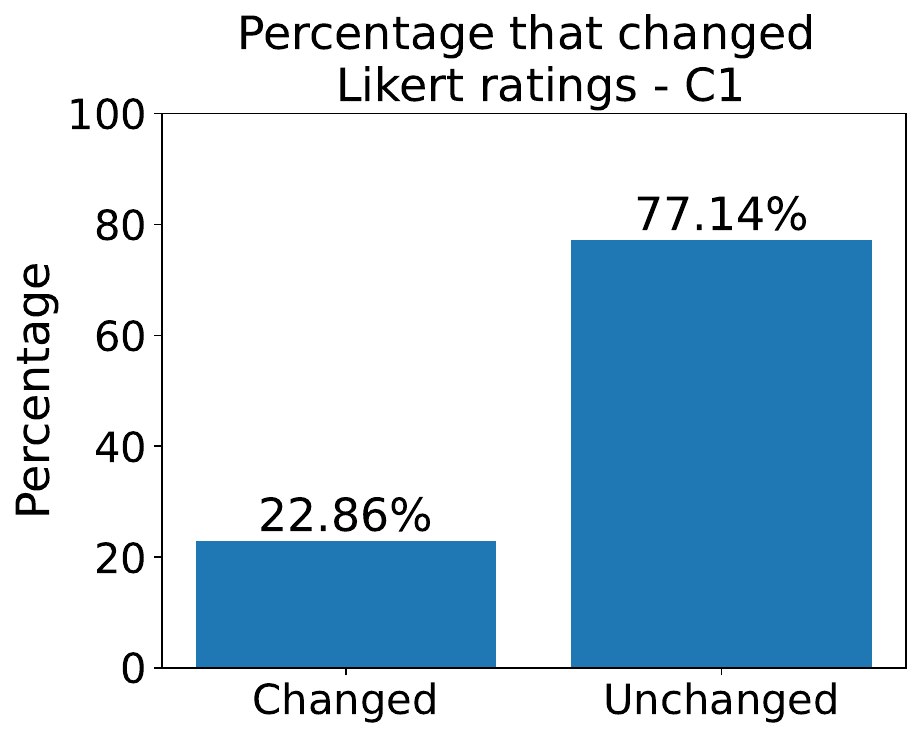}
  \label{fig:sub1}
\end{subfigure}%
\begin{subfigure}
  \centering
  \includegraphics[width=.35\linewidth]{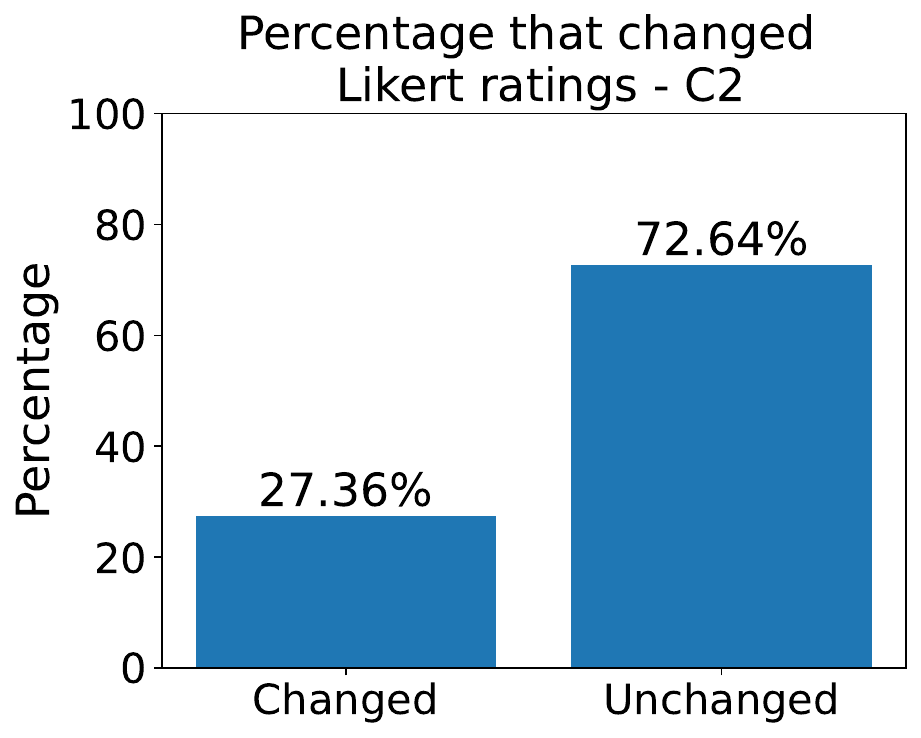}
  \label{fig:sub2}
\end{subfigure}
\caption{\small{(a) Percentage of times participants changed Likert ratings in Condition 1 (similar condition). Participants changed Likert ratings 22.86 percent compared to not changing 77.14 percent of the time. (b) Percentage of times participants changed Likert ratings in Condition 2 (randomized condition). Participants changed Likert ratings 27.36 percent time compared to not changing 72.64 percent of the time.}}
\label{fig:percentage_change}
\end{figure*}

\begin{figure*}
\centering
\begin{subfigure}
  \centering
  \includegraphics[width=.35\columnwidth]{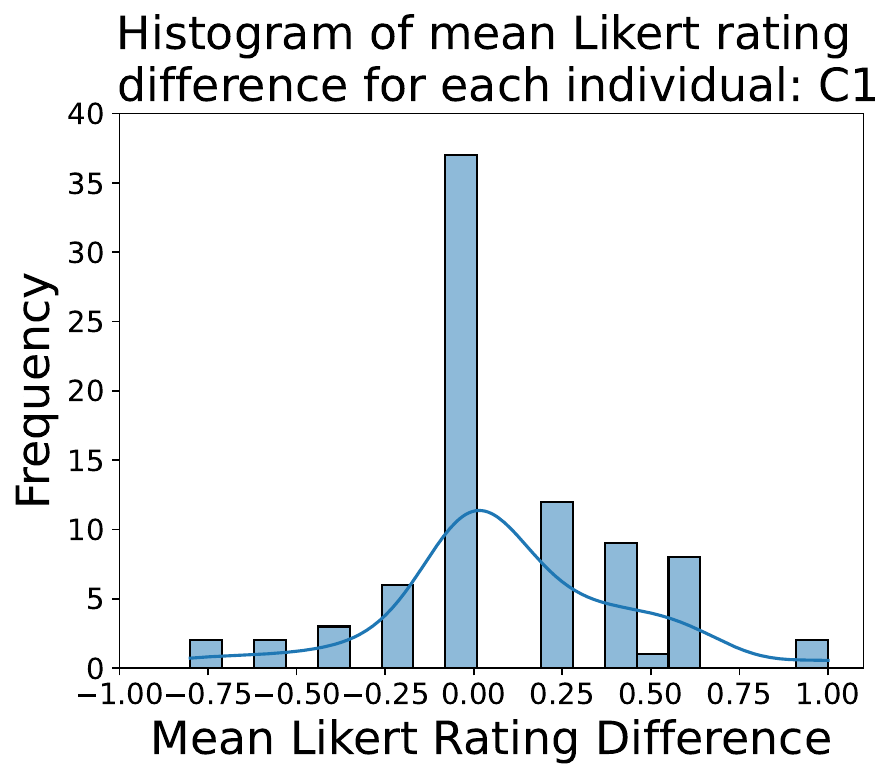}
  \label{fig:sub1}
\end{subfigure}%
\begin{subfigure}
  \centering
  \includegraphics[width=.35\columnwidth]{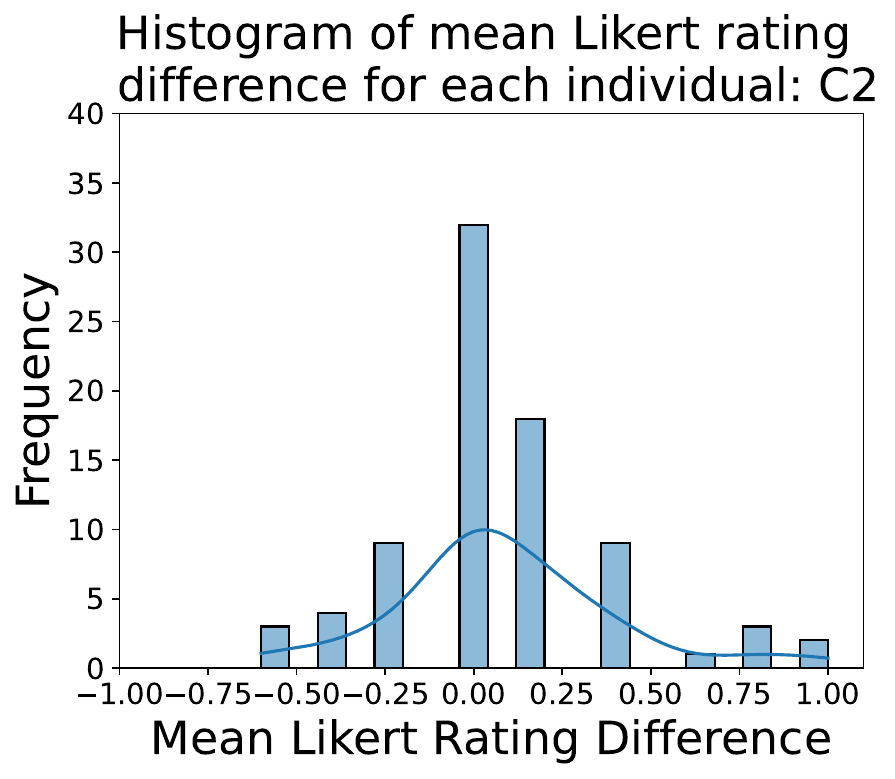}
  \label{fig:sub2}
\end{subfigure}
\caption{\small{(a) Histogram of mean change in Likert rating for each individual in Condition 1 (similar condition) (b) Histogram of mean change in Likert rating for each individual in Condition 2 (randomized condition) In both cases, our results show that individuals are mostly rigid in their beliefs and do not drastically change them.}}
\label{fig: per_person_likert_diff}
\end{figure*}

Aggregating across both conditions, our analysis shows that participants update their beliefs 25.13 percent of the time (C1 = 22.86 percent, C2 = 27.36 percent; no significant difference between the two conditions according to two-sample z-test) (Figure \ref{fig:percentage_change}), showing that beliefs are rigid for most people (Figure \ref{fig: per_person_likert_diff}). Despite this rigidity, peer responses in stage 2 affect individual shifts in belief, consistent with existing literature that suggests that beliefs are indeed influenced by peer beliefs \cite{coppock2023persuasion}. Our analysis shows that individual belief changes to be more similar to the beliefs of their peers. We define $\delta R_{i}$ as the difference between the participant's updated Likert rating in stage 2 and the initial Likert rating in stage 1 for a specific participant, $i$. Moreover, we define social signal, $S_{i}$ as the average rating of the peers in stage 2, and $\delta S_{i}$ as the difference between $S_{i}$ and the initial Likert rating of the participant. We find positive correlation between $\delta S_{i}$ and $\delta R_{i}$, showing that peers can influence individual beliefs. For similar condition group (C1), $\rho$ is 0.338 (p: 2.378e-11) [CI: 0.244, 0.425], and for randomized condition group (C2),  $\rho$ is 0.310 (p: 2.140e-10) [CI: 0.219, 0.396]. Note that in C2, we intentionally show participants response entries they did not select for themselves. Yet, the $\rho$ is similar across both conditions, suggesting that people are likely to be influenced by their peers' opinions, regardless of whether these opinions align with their own or not. A stronger indication of this phenomenon is seen when we consider only instances where participants did change their responses. In such instances, we observe a significant increase in correlation for both conditions. For C1, $\rho$ is 0.624 (p: 2.378e-11)[CI:0.475, 0.738], and for C2, $\rho$ is 0.622 (p: 2.140e-10) [CI: 0.493, 0.725]. 


However, individuals often do not realize the impact of peer influence on their beliefs. In stage 2, participants select from a predetermined list the reason behind the change in their stance (or the lack thereof). Only a small percentage selects that peer responses influenced their change in rating (13.25\% in C1, and 16.58\% in C2) compared to the percentage that actually changed their rating (no statistical difference between the proportions in the two conditions as per proportion z-test). When considering only participants that did change their ratings, we see that only about half attributed that change to peer influence (47.25\% in C1, 50.91\% in C2; no statistical difference between the proportions as determined by proportion z-test). Interestingly, participants in Condition 1 (C1) reported not changing their minds a higher percentage of times than those in Condition 2 (C2). For C1, 83.90\% of the responses included ``I did not change my mind", while for C2, the percentage was 75.88\% (p-value of proportion z-test: 0.005). We can interpret this analysis as some indication that the randomized condition group (C2), where participants are presented with a wider range of viewpoints, is more ``aware" of their belief shifting. This insight is particularly compelling when we also consider that there was no significant difference between the two conditions regarding how often participants change their responses.   

\begin{figure*}
\centering
\begin{subfigure}
  \centering
  \includegraphics[width=.35\columnwidth]{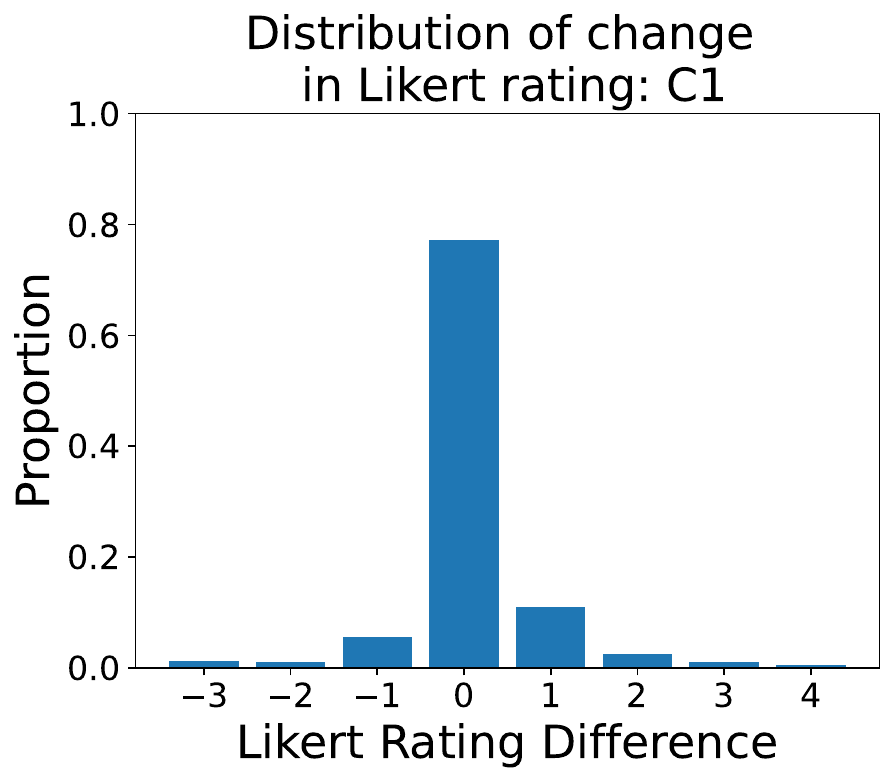}
  \label{fig:sub1}
\end{subfigure}%
\begin{subfigure}
  \centering
  \includegraphics[width=.35\columnwidth]{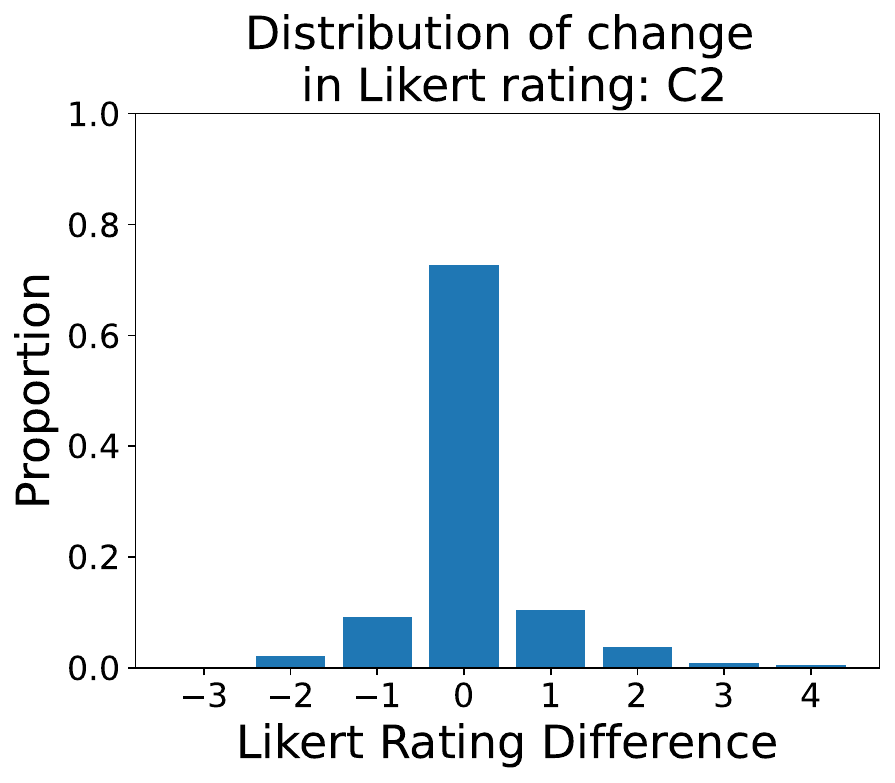}
  \label{fig:sub2}
\end{subfigure}
\caption{ \small{(a) Distribution of mean change in Likert rating in Condition 1 (similar condition) (b) Distribution of mean change in Likert rating in Condition 2 (randomized condition) Both distributions peak at 0 and are unimodal.}}
\label{fig: likert_diff_dist}
\end{figure*}

It must be noted that most individuals do not drastically change their beliefs. Instead, the shift in belief is more gradual, consistent with previous literature \cite{scheffer2022belief, sharot2011unrealistic}. Plotting the change in Likert scale for every instance, we see an unimodal distribution, peaking at 0 (Figure \ref{fig: likert_diff_dist}), with more extreme values being less common. Note that the scale never reaches 6 or -6, also showing that the shift in belief is not drastic. There is no significant difference between the distribution of change in Likert rating between the two conditions (as shown by Levene's Test of variance). The same phenomenon is seen when considering the distribution of the average change in Likert rating for each participant (Figure \ref{fig: per_person_likert_diff}) as the distribution peaks at 0. 

In summary, our results suggest that despite individual beliefs being rigid, peer opinions can influence individuals, irrespective of what the opinion entails, providing empirical evidence for RQ1. 



\subsection*{Introducing randomness while recommending can increase incorporation of diverse views into individual's networks} 
\begin{figure*}
\centering
\begin{subfigure}
  \centering
  \includegraphics[width=.35\linewidth]{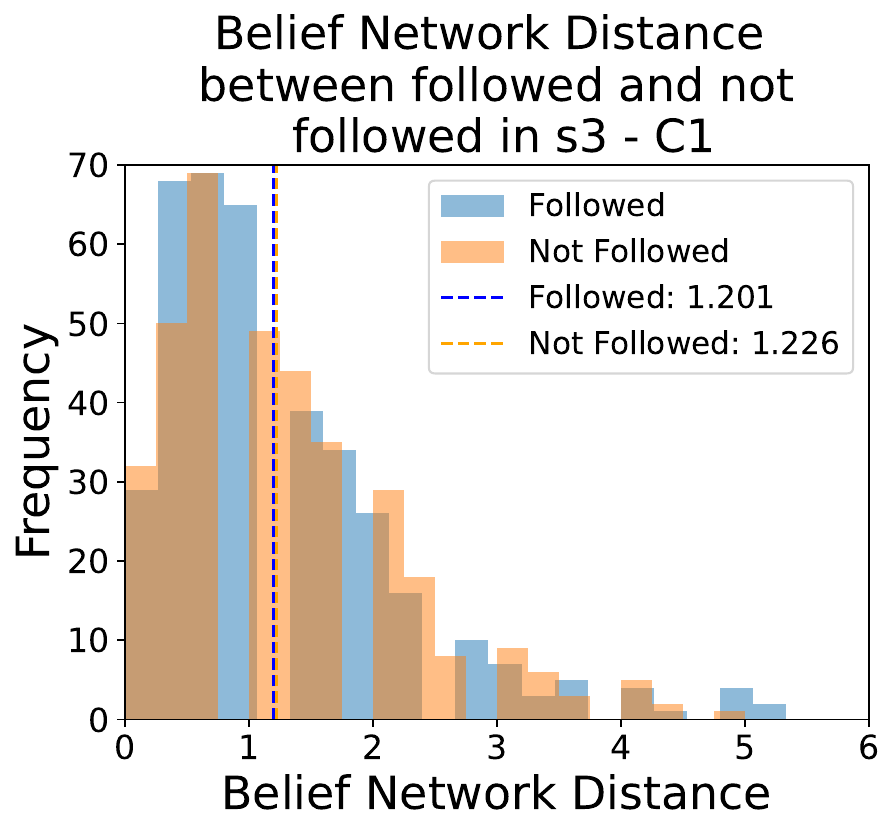}
  \label{fig:sub1}
\end{subfigure} 
\begin{subfigure}
  \centering
  \includegraphics[width=.35\linewidth]{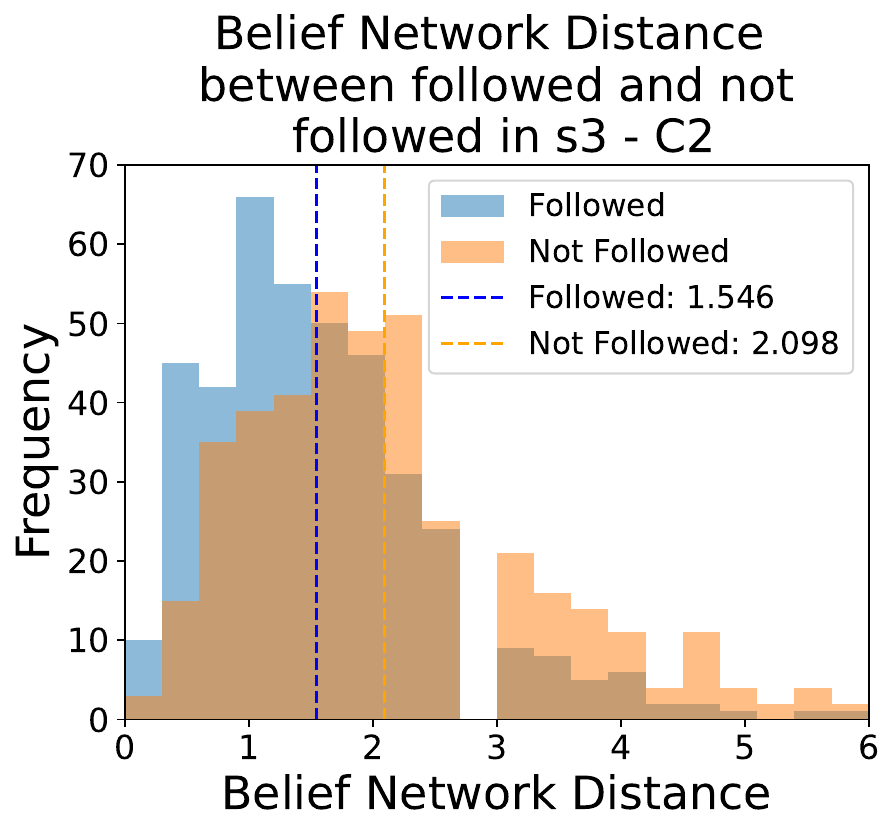}
  \label{fig:sub2}
\end{subfigure}
\caption{\small{(a) Belief Network Distance for Condition 1 (similar condition). Mean Followed (denoted by blue line) is not statistically different from mean Not followed (denoted by orange line). (b) Belief Network Distance for Condition 2 (randomized condition). Mean Followed (denoted by blue line) is lower than mean Not followed (denoted by orange line), signifying people's homophilic tendencies. However, mean Followed in C2 is greater than mean Followed in C1, showing people's tendency to incorporate a wider variety of beliefs to some extent.}} 
\label{fig: network_belief_distance}
\end{figure*}

Our findings indicate that participants tend to choose responses that align with their own, meaning they usually follow others who share similar opinions. This tendency is particularly noticeable in the randomization condition, where a wide variety of opinions are presented. We calculate the Belief Network Distance (defined in Equation \ref{eq:B}) for both those followed in stage 3 ($B_{i_{followed}}$, see Equation \ref{eq:B_followed}) and those not followed but shown in stage 3 ($B_{i_{notfollowed}}$, see Equation \ref{eq:B_notfollowed}) and compare them. The Belief Network Distance signifies how far the average belief of the peers in the user's network for that round is from the participant's belief. Denoting $B_{i}$ as the Belief Network Distance for each participant $i$, mean $B_{i_{followed}}$ is 1.201 and mean $B_{i_{notfollowed}}$ is 1.226 for C1. There is no significant difference between the two (using t-test), which makes sense since C1 is designed to only show similar content. For C2, mean $B_{i_{followed}}$ is 1.546 and mean $B_{i_{notfollowed}}$ is 2.098. Our t-test shows a significant difference between the two (p: 1.733e-12), implying that even when presented with a diverse set of views, participants are more likely to select those similar to their own (Figure \ref{fig: network_belief_distance}). 

Moreover, our results from calculating the correlation between the follow signal, $F_{i}$ (defined in Equation \ref{eq:F_i}), and the participant's belief show the same phenomenon - there is a moderately strong positive correlation between a participant's belief and the beliefs of who they follow. When calculating $\rho$ between $F_{i}$ and the initial Likert rating of the participant, we find $\rho$ to be 0.645 (p: 4.897e-46) [CI: 0.582, 0.700], and for C2, $\rho$ is 0.503 (p: 3.457e-27) [CI: 0.426, 0.573]. Additionally, we consider the updated Likert rating to be the participant's current belief state, and we calculate the correlation between follow signal, $F_{i}$, and the participant's updated Likert rating.  For C1, $\rho$ is 0.636 (p: 9.744e-45) [CI: 0.572, 0.692], and for C2, $\rho$ is 0.519 (p: 2.659e-29) [CI: 0.444, 0.587].   

\begin{figure*}
\centering
\begin{subfigure}
  \centering
  \includegraphics[width=.35\linewidth]{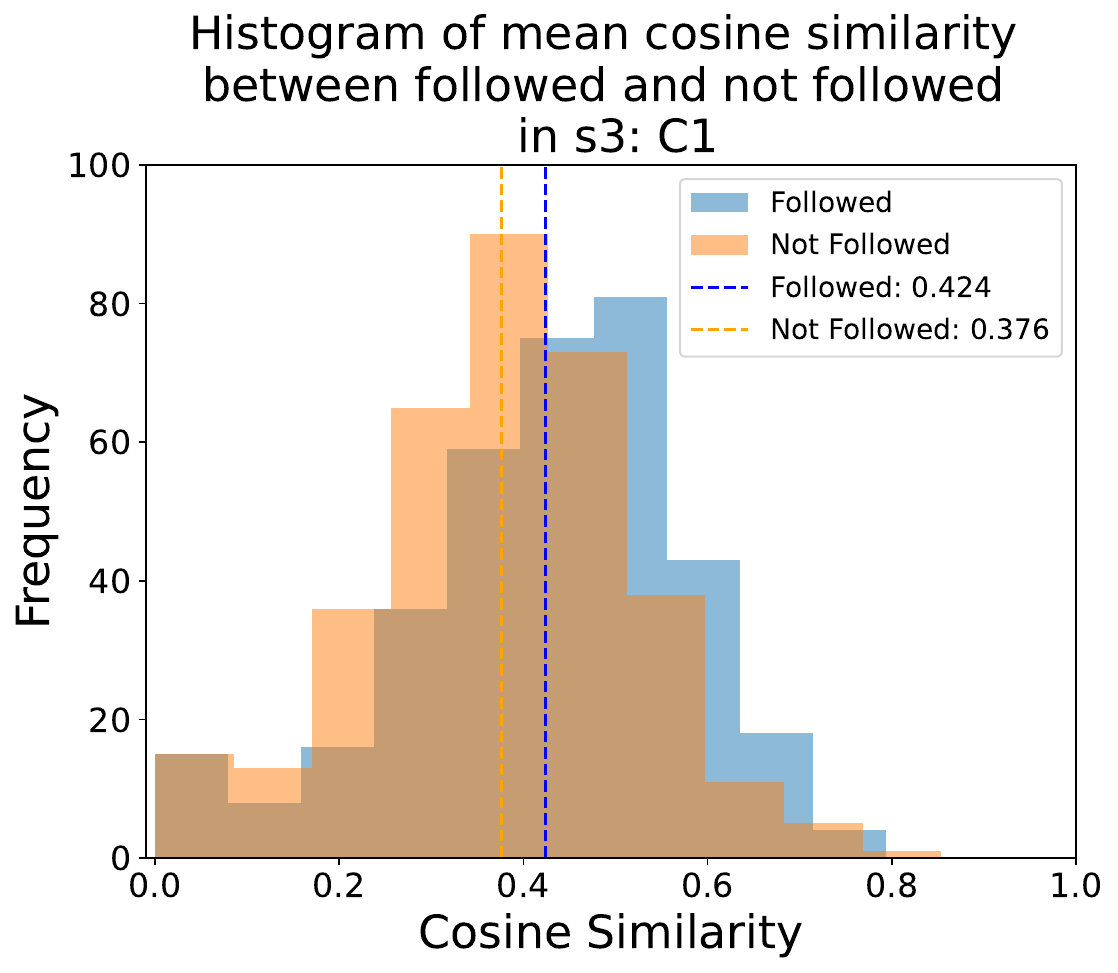}
  \label{fig:sub1}
\end{subfigure}%
\begin{subfigure}
  \centering
  \includegraphics[width=.35\linewidth]{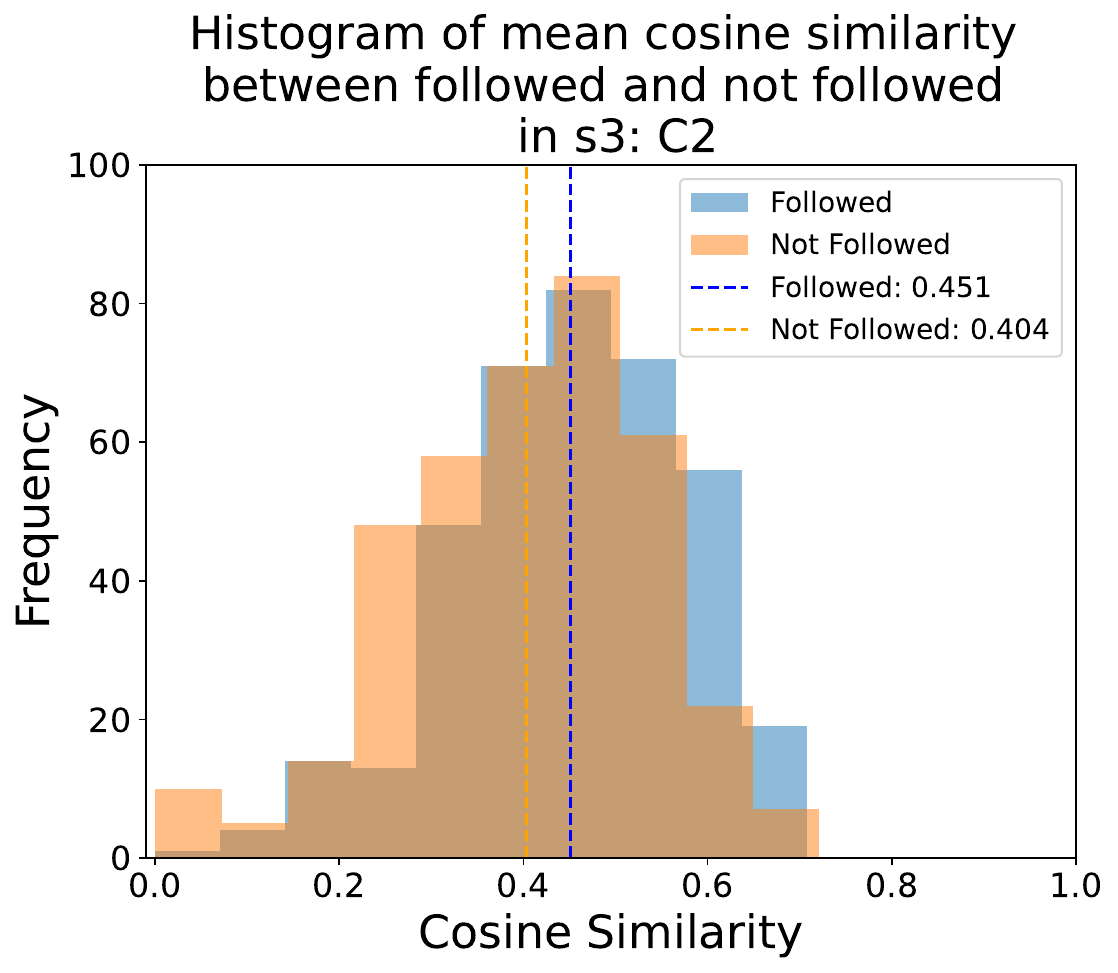}
  \label{fig:sub2}
\end{subfigure}
\caption{ \small{(a) Comparing the distribution of cosine similarity for followed and not followed in stage 3 for Condition 1 (similar condition) (b) Comparing the distribution of cosine similarity for followed and not followed in stage 3 for Condition 2 (randomized condition). Mean followed (denoted by blue line) for both conditions are semantically more similar to the participant compared to those not followed (denoted by orange line).}}
\label{fig:cosine_similarity_third_stage}
\end{figure*}

Furthermore, we evaluate the cosine similarity between an individual's reasoning and the reasoning of the responses they follow, compared to the cosine similarity between an individual's reasoning and the reasoning of those they do not follow (Figure \ref{fig:cosine_similarity_third_stage}). Our results show that the response entries followed by participants are semantically more similar to their own answers compared to those not followed in both conditions. For C1, the mean cosine similarity for followed is 0.424; for not followed, the mean cosine similarity is 0.376 (significant difference using Mann-Whitney U test, p: 2.473e-05). For C2, the mean cosine similarity for followed is 0.451, and for not followed, it is 0.404 (significant difference using Mann-Whitney U test, p: 9.661e-07). 

However, it is possible to reduce some of the homophilic tendency of individuals by increasing randomness in the recommendation algorithm and presenting individuals with a diverse set of views. Mean $B_{i_{followed}}$ of C1 is less than the mean $B_{i_{followed}}$ of C2 (using t-test, p: 1.305e-06), showing that by increasing diversity in the content of response entries, it is possible to increase being open to incorporating other beliefs. Similarly, the correlation between the follow signal, $F_{i}$, and the participant's belief is lower in the randomized condition (C2) compared to the similar condition (C1) ($\rho$ mentioned above). Our results imply that people tend to incorporate other beliefs to \textit{some} extent when they have access to them. Yet, we do not find any semantic difference in the response reasoning between C1 and C2 for followed and not followed. Despite a higher average cosine similarity in C2 than C1 for both followed and not followed, no statistical differences were found between the two conditions (Mann-Whitney U Test, p: 0.012 for followed; p: 0.009 for not followed).

Summarizing our results to answer RQ2, we find that algorithmic changes in recommendation method, i.e., showing a wider diversity of views instead of opting for homophily-based recommendation, can reduce homophilic tendencies, albeit slightly. Our results also show that people are willing to incorporate more diverse opinions into their networks when they are exposed to them. 

\section{Discussion}

Existing research indicates that online social networks play a critical role in influencing individuals' political attitudes, behaviors, and perceptions \cite{asimovic2021testing, valenzuela2012social}. Our findings reiterate this since regardless of whether peer opinions align or vary, we see that they impact individuals' beliefs. This also implies that traditional social networks, which promote similar content, can indeed lead to increased belief rigidity, potentially exacerbating polarization. Our research empirically shows that a simple tweak such as introducing some randomness in what is displayed to users can have a positive impact on belief rigidity. In other words, when given the option and access to a wider range of viewpoints, people are inclined to include at least some of those viewpoints in their networks. This provides empirical evidence that making social networks less tailored to individual preferences could potentially encourage openness to different viewpoints.



In this paper, we take an empirical approach by building an experimental framework, which we use to understand belief rigidity on online social networks. In the first stage, participants are given a prompt to rate, and in the second stage, they can change their rating based on the responses of their peers. This technique is similar to a Delphi study, which has been used in various experimental frameworks to understand individual and collective attitudes and beliefs \cite{moussaid2013social}. These two stages allow us to understand the impact of peers on individual beliefs. Our third stage is designed to quantify the impact of recommendation algorithms, allowing participants scope for network restructuring. Therefore, our experimental framework enables us to take into account the combined effect of human social influence and algorithmic decisions, consistent with research on opinion dynamics in online social networks \cite{santos2021link}. 

While social media analysis is generally used to study social phenomena, it is difficult to test out different intervention strategies and compare them in real time on social media platforms. Moreover, in real social platforms individuals not only view each other's posts but also directly interact with them. Social media analysis can therefore only estimate the change in belief since online behavior may differ from people's private beliefs due to the online disinhibition effect \cite{suler2004online, lieberman2020two}. For example, people may appear more stubborn or even aggressive just to win an online argument \cite{suler2004online}. In our experimental framework, there is no peer pressure (since participants are anonymized) or fear of social isolation (since participants don't have access to how many ``followers'' they have) compared to actual social media platforms. Since participants have no external motivation to change beliefs, our experimental framework enables passively sensing \textit{private belief}. 

Our results show that individuals' beliefs are influenced by peer opinions even when the peer opinions are different from their own. This has been theoretically established in existing literature for opinion dynamics in social networks - dissimilar peer opinions tend to moderate individual opinions \cite{santos2021link}. However, there is often a lack of realization regarding this influence among participants. Change in belief is often minor, suggesting that beliefs are inherently rigid but can be gradually influenced over time. This is consistent with Introne's proposed Belief Landscape Framework, where he showed that people who update their beliefs tend to do so incrementally as if moving through a landscape composed of stable regions (where beliefs change slowly) and transitional regions (where beliefs change more quickly) \cite{introne2023measuring}. Tying our findings back to our first research question, it is, therefore, important to diversify exposure to different viewpoints to facilitate slight but meaningful shifts in beliefs within social networks. 

We also observe the homophilic tendency of individuals to prefer and follow content that align with their own beliefs, consistent with existing literature \cite{lewis2008tastes, itkonen2015social, bessi2016homophily}. For example, a study on climate change link-sharing on Facebook shows that users are mostly online friends with others who share similar concerns about the topic \cite{itkonen2015social}. However, we show that by introducing randomness in recommendation algorithms, this homophilic tendency can be slightly reduced, encouraging openness to differing views. Similar findings have been seen in other domains such political polarization --  exposure to counter-attitudinal news (news that contradicts one's beliefs) on Facebook was shown to reduce affective polarization \cite{levy2021social}. Tying this finding to our second research question, a randomized recommendation can convince people to diversify peer choices, but not drastically. 

However, our study still has some limitations. One limitation is that this study is population-dependent since the results of each experiment would strongly depend on the demographic make-up of that batch. Our sample size is moderate (N = 163) and we only considered participants from the US. While that might reduce the generalizability of our findings, it must be noted that the whole experiment is synchronous, takes an hour, and all participants must join at the same time. Therefore, our experiment requires a significant time and resource commitment  (needing access to a computer with an internet connection for an hour at a provided time) from participants. Previous literature has discussed the difficulties of recruiting participants for relatively longer complex experiments. Levendusky et al. identify that it is difficult to recruit participants for studies over twenty minutes even with relatively larger incentives, especially when relying on online survey takers \cite{levendusky2021we}. However, complex experiments can provide a more in-depth understanding. Therefore, for complex studies, a smaller N is often an acceptable trade-off. Some examples of such studies include understanding idea generation in social networks \cite{baten2022novel}, facilitating eco-friendly purchases through user interface design \cite{IslamSEER}, or designing technology to reduce misinformation \cite{jahanbakhsh2024browser} and so on. This can be considered a ``depth–breadth trade-off'', where the researcher has to balance between the complexity of the task and the number of participants \cite{levendusky2021we}. For our study, we therefore chose a more complex setup and a smaller sample size. 


While our framework has some features present in traditional social media platforms, the framework itself does not imitate a traditional social media platform. Our framework has three stages in every round.  Participants provide a response to a prompt in stage 1, see their peer's responses in stage 2, and follow/unfollow people in stage 3. In social platforms, though, people can share content, interact with others, and follow and unfollow content all at once. We acknowledge that the interactions on social media platforms are far more complex. However, breaking these interactions down to simpler forms as we did here allows us to probe into the individual impact of each of these features. In future studies, our framework can be used to compare different design choices or different algorithmic choices for different platform components. 

Similarly, in our study, all participants were anonymized, did not have any identifiable cues, and were unknown to the other participants. While there are some social media platforms where people have very little information about others (such as Reddit), that is not the case for most platforms such as Facebook, Instagram, and X (formerly Twitter). Platforms often display user identity, follower count, or various demographic cues, which significantly influence trust and perception \cite{suh2010want, abrahao2017reputation}. So, it is crucial to study the impact of displaying demographic cues and follower count on belief rigidity. Our framework facilitates designing such studies since minimal changes would be required to include displaying cues in the study. Likewise, the statement prompts can also be changed to adapt our study to other topics such as health, vaccines, or gun control. 

Finally, our study aims to measure immediate changes in belief rigidity. According to the Belief Landscape framework \cite{introne2023measuring}, shifts in belief are often slight and subtle, which is what we observed in our study. Still, it would also be interesting to conduct a longitudinal study to quantify the long-term effects of exposure to diverse viewpoints on belief change. A longitudinal experiment would allow us to map the ``tipping points'' where people change their beliefs more drastically. 

\section{Conclusion}
In this paper, we explore whether a more randomized recommendation can convince people to diversify peer choices, and aim to quantify the impact of this diversification. Our results show that increasing randomization in a network can improve people's ability to incorporate a higher diversity of views, despite people's natural homophilic tendency. Moreover, we show that individuals are affected by the opinions of others, irrespective of if they hold the same opinions as them. While we acknowledge that beliefs are inherently rigid, our study shows that there are plausible steps to be taken to reduce belief rigidity. Our experimental framework can easily be modified to study different belief rigidity cues such as trust, reputation, and popularity, and we encourage others to do so. Finally, although our work is empirical, we can use this method to draw out meaningful connections in the real world. 

\section*{Materials and Methods}
Here, we provide details about the experimental framework design, study setup, validation checks for our setup, and participant demographic information. The experimental framework is shown in Figure~\ref{fig:experimentalCondition}. 

\subsection*{Experimental Framework Design}
Our study consists of five rounds. In each of the rounds, participants are given one prompt regarding climate change policies, and the focus of the round is on that statement. The prompts are designed to probe the stance of people regarding spending money to mitigate climate change and are derived from the Climate Change Twitter Dataset \cite{climatechangeTwitter}, which is a Twitter dataset containing actual posts on climate change on Twitter, and ClimateFever \cite{climatefever}, which is a fact-checking dataset on climate change. Our criteria for the prompts was that they must be short and unambiguous. For both conditions, the prompts are the same (further details in Supplementary materials).

Each round consists of three stages - response, revision, and selection. Additional information on interface design is provided in Supplemental materials. The stages are explained below. 

\subsubsection*{Stage 1: Response stage} 
Given a prompt, participants are asked to evaluate each statement using a 7-point Likert scale, where 0 represents ``strongly disagree" and 6 signifies ``strongly agree". They must also explain the reasoning for their rating in a maximum of 400 characters. We refer to each evaluation and its accompanying explanation as a ``response entry''. 

\subsubsection*{Stage 2: Revision stage} 
In the revision stage, participants are shown the response entries of a selected group of other participants (referred to as ``peers"). How this group of peers is selected depends on the round and whether the participant is part of the similar condition or the randomized condition. 

\textbf{Condition 1:} For the first round in C1, each participant is matched with three peers — referred to as their base connections — based on the similarity of their responses to the first prompt. Essentially, this means that the peers gave the prompt similar ratings as the participant. For the subsequent rounds, participants see the response entries of peers they selected in stage 3 (selection stage) of the immediate previous round. 

\textbf{Condition 2:} For the first round in C2, participants are randomly assigned three base connections. In subsequent rounds, participants still see three peers. However, unlike C1, only one of these peers is chosen by the participant themselves during stage 3 (selection stage) of the previous round. The other two peers are chosen randomly from among those the participant did not follow.

As the participants have access to the response entries of a selected group of peers, participants are now asked to rate the same prompt again on a 7-point Likert scale. To ensure their change in rating is intentional, we display the participant's initial Likert rating to the prompt. However, they still need to click the Likert scale again manually. This difference between the updated Likert rating and the initial Likert rating can be used to represent belief rigidity. Note that, belief rigidity is defined as the ability to change beliefs in light of new information/opinion, and the peer response entries can be considered as the ``new information/opinion". 

\subsubsection*{Stage 3: Selection stage} 
In the selection stage, participants are presented with the response entries of six participants in total - including the three peers they see in stage 2 (and the rest are ``recommended" to them). The recommendation process depends on the round and the group the participant is part of. For the first round, the three participants recommended are those similar to the participant for C1, and for C2, they are randomly selected from the pool of other participants (excluding the base connections).  


The participants must select (i.e., ``follow") at most three participants whose answers they would like to see in stage 2 (revision stage) for the next round. They can stick to the connections assigned or pick new ones (or do a mix of both). However, as mentioned previously, only C1 participants see who they followed in stage 2. C2 participants get only one of the followed peers in stage 2, and the other two are randomly selected from those not followed. The participants must also like or dislike all the response entries shown. 

For the upcoming rounds, the recommendation process in stage 3 is designed in the following way - 
\begin{itemize}
    \item \textbf{Condition 1:} Participants see the people they are currently following, and three other people whose responses are the most similar to them.  
    \item \textbf{Condition 2:} Participants see the three people they are currently following, the two people randomly selected and shown to them in stage 2, and another randomly selected participant. 
\end{itemize}

Once again, participants must select (``follow") three participants whose answers they would like to see in stage 2 (revision stage) of the subsequent round, and the process repeats for the remaining rounds. An example of the user experience flow is presented in Supplementary materials.

\subsection*{Study Details} 
The experimental platform was designed using Empirica v1 \cite{almaatouq2021empirica}, a Javascript framework designed for facilitating multi-user experiments in real time. Participants were recruited through Prolific Academia, a crowd-sourcing platform aimed at connecting research participants with behavioral studies \cite{peer2017beyond}. The criteria for inclusion were that participants must be located in the US, have access to a computer and internet connection, and must be able to join the study at the time being conducted. The study lasted for approximately 60 minutes, and participants were given \$15 after completion of the study. Participants provided informed consent while signing up for the study, and again while starting the study. There were 5 batches of C1 and 5 batches of C2. 

\begin{figure*}
\centering
\begin{subfigure}
  \centering
  \includegraphics[width=.35\linewidth]{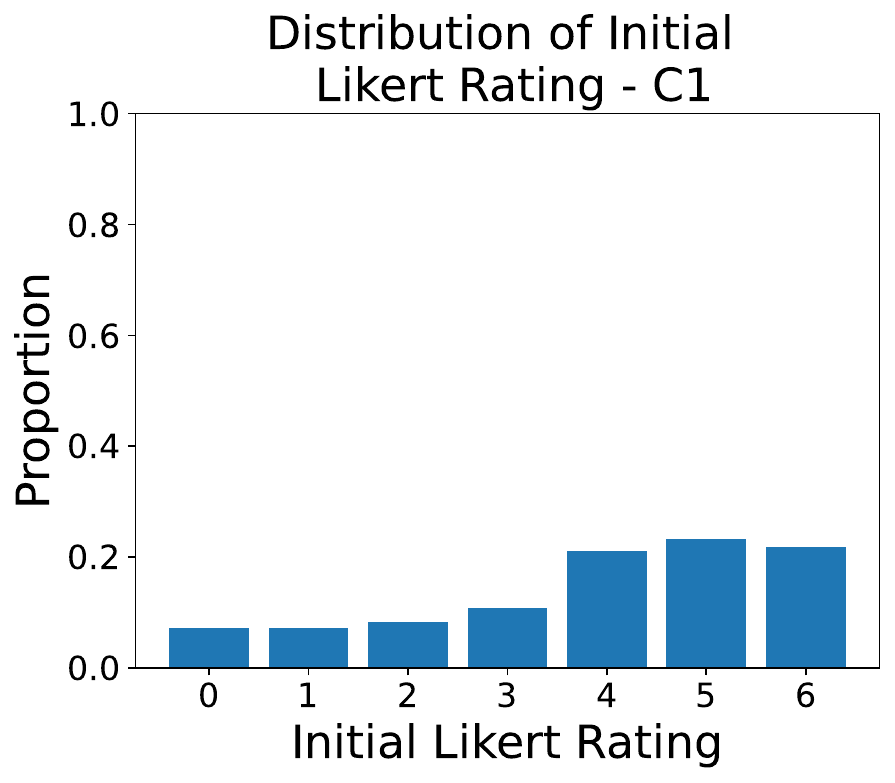}
  \label{fig:sub1}
\end{subfigure}%
\begin{subfigure}
  \centering
  \includegraphics[width=.35\linewidth]{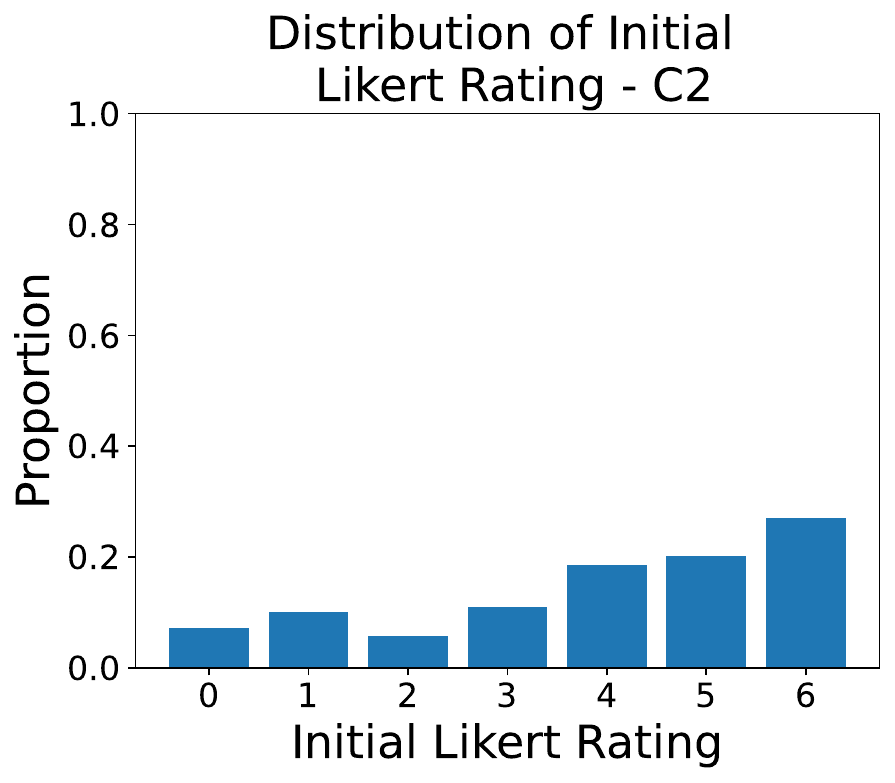}
  \label{fig:sub2}
\end{subfigure}
\caption{\small{(a) Distribution of initial Likert rating for Condition 1 (similar condition) (b) Distribution of initial Likert rating for Condition 2 (randomized condition). The distribution suggests that more participants rated agreeably to the statements, which is representative of the current US sentiment regarding climate change.}}
\label{fig: initial_likert_dist}
\end{figure*}

Aggregating across all batches, C1 had 82 participants and C2 had 81 participants (total N = 163). The participant demographic is added to Supplementary materials. In general, there were similar proportions of gender and race in both conditions. Most participants showed pro-climate sentiment, with the aggregate average initial Likert response being 3.91 (C1: 3.89, C2: 3.93) and aggregate skewness being -0.680 (C1: -0.696, C2: -0.669) (Figure \ref{fig: initial_likert_dist}). This means that the distribution is moderately left-skewed, which indicates that more participants had an agreeable response to our statements. This is representative of the current sentiment in the US, where the majority of the US population shows pro-climate sentiment \cite{yaleClimateChange}.

\begin{table}[t!]
\centering
\caption{Definition of the symbols used in this paper}
\begin{tabular}{ll}
Symbol & Parameter\\
\midrule
R & Rating of the Likert scale \\
i & Participant \\
j & Peer \\
S & Social signal  \\
F & Follow signal \\
B & Belief Network Distance \\
G & Number of participants being followed \\
\bottomrule
\end{tabular}
\end{table}

\subsection*{Defining Analysis Measures}

\subsubsection*{Belief Update}
We define the change in belief as the difference between the initial rating given by the participant in stage 1 and the updated rating of the participant in stage 2. Assuming $R$ is the rating, and $i$ denotes each participant, let $R_{i_{initial}}$ and $R_{i_{updated}}$ denote the initial and updated ratings for the $i^{th}$ participant, respectively. The change in belief for each participant, $\delta R_i$, can then be represented as:

\begin{equation}
\label{eq:R_i}
\delta {{R}_{i}}= R_{i_{updated}} - R_{i_{initial}}
\end{equation}

\subsubsection*{Social Signal}
We define social signal, $S$, as what the participant sees in the revision stage, and this gives the participant some direction to what peers in the network might think about the same topic. To calculate social signal for each participant for a particular round - $S_{i}$ - we take the mean of the ratings of peers in the revision stage. Assuming $j$ denotes each peer, and $N$ is the total number of peers,

\begin{equation}
\label{eq:S_i}
     S_{i} = \frac{\sum_{j=1}^{j=N} R_{j_{initial}} }{N}
\end{equation}

To calculate how different the social signal, $S_{i}$ is from their own rating, we subtract the initial rating of the participant, $R_{i_{initial}}$ from the social signal, $S_{i}$. Therefore, $\delta S_{i}$ is defined as 

\begin{equation}
\label{eq:delta_S_i}
\delta S_{i} = S_{i} - R_{i_{initial}}
\end{equation}

Intuitively, $\delta S_{i}$ is measuring how different peer responses are from the individual.  


\subsubsection*{Follow Signal}
For a particular round, for each participant, we define follow signal, $F$, as the ratings of those the participant chooses to follow in the selection stage. To calculate the follow signal for each participant for a specific round - $F_{i}$ - we take the mean of the ratings of the followed peers. Assuming $j$ denotes each followed peer, and $N$ is the total number of followed peers,

\begin{equation}
\label{eq:F_i}
    F_{i}= \frac{\sum_{j=1}^{j=N} R_{j_{initial}}}{N}
\end{equation}



\subsubsection*{Belief Network Distance}
Extending the concept of follow signal, $F_{i}$, we define belief network distance, $B_{i}$ as the average difference between a participant's updated rating and the initial rating of other participants within a belief network. This signifies how different the belief is compared to the participants' own beliefs. We consider the updated Likert rating for the participant since that is their most current belief, but we consider the initial belief of others since that is what they see as part of the response entries. Assuming $N$ is the total number of participants shown for that given round,
\begin{equation}
\label{eq:B}
     B_{i} = \frac{\sum_{j=1}^{j=N} |R_{j_{initial}} - R_{i_{updated}}|}{N}
\end{equation}

Modifying the above equation to include only those that the participant is following, we can denote $G$ as followed. 

\begin{equation}
\label{eq:B_followed}
     B_{i_{followed}} = \frac{\sum_{j=1}^{j=G} |R_{j_{initial}} - R_{i_{updated}}|}{G}
\end{equation}

Denoting participants that are not followed as $G'$, the equation becomes 

\begin{equation}
\label{eq:B_notfollowed}
     B_{i_{notfollowed}} = \frac{\sum_{j=1}^{j=G'} |R_{j_{initial}} - R_{i_{updated}}|}{G'}
\end{equation}

\subsection*{Validating the study setup}
To ensure that C1 shows similar signals as the participant's own choice and C2 shows more diverse responses, we calculate the Pearson correlation, $\rho$, between the participant's initial Likert rating, and the average of the peer rating shown in stage 2. The $\rho$ for C1 is 0.424 (p: 1.309e-17) [CI: 0.337, 0.504], and the $\rho$ for C2 is 0.178 (p: 3.313e-04) [CI: 0.082, 0.271], implying that C1 participants indeed see more similar peer responses compared to the participants in C2. To ensure that the recommendation algorithms are significantly different between the two conditions, we calculate the Pearson correlation, $\rho$, between the mean of all Likert ratings displayed in stage 3 with the initial Likert rating of the participant. For C1, $\rho$ is 0.747 ( p: 1.865e-63) [CI: 0.696, 0.790] and for C2, $\rho$ is 0.311 ( p: 2.563e-10) [CI: 0.219, 0.398].  The correlations show that the recommendations provided for C1 are more similar to the participant compared to the recommendations provided for C2, which was the intention of the design. Moreover, the average number of words written as a response reason for C1 is $23.48$ (median: $21$) and for C2 is $28.04$ (median: $25$). This shows that participants elaborated on their opinions, increasing the trustworthiness of the framework.

\bibliographystyle{ieeetr}
\bibliography{ref}  

\end{document}